\documentclass[preprints,article,accept,moreauthors,pdftex]{Definitions/mdpi} 

\firstpage{1} 
\pubvolume{1}
\issuenum{1}
\articlenumber{0}
\pubyear{2021}
\copyrightyear{2020}
\datereceived{} 
\dateaccepted{} 
\datepublished{} 
\hreflink{https://doi.org/} 

\Title{Estimating the Parameters of Extended Gravity Theories with the Schwarzschild Precession of S2 Star}

\TitleCitation{Estimating the Parameters of Extended Gravity Theories with the Schwarzschild Precession of S2 Star}

\Author{Du\v{s}ko Borka$^{1,*}$\orcidA{}, Vesna Borka Jovanovi\'{c}$^{1}$\orcidB{}, Salvatore Capozziello$^{2,3}$\orcidC{}, Alexander F. Zakharov$^{4,5}$\orcidD{} and Predrag Jovanovi\'{c}$^{6}$\orcidE{}}

\AuthorNames{Du\v{s}ko Borka, Vesna Borka Jovanovi\'{c}, Salvatore Capozziello, Alexander F. Zakharov and Predrag Jovanovi\'{c}}

\AuthorCitation{Borka, D.; Borka Jovanovi\'{c}, V., Capozziello, S.; Zakharov, A. F.; Jovanovi\'{c}, P.}

\address{%
$^{1}$ \quad Department of Theoretical Physics and Condensed Matter Physics (020), Vin\v{c}a Institute of Nuclear Sciences - National Institute of the Republic of Serbia, University of Belgrade, P.O. Box 522, 11001 Belgrade, Serbia \\
$^{2}$ \quad Dipartimento di Fisica " E. Pancini", Universit\`a  di Napoli "Federico II" and Istituto Nazionale di Fisica Nucleare (INFN) Sez. di Napoli, Compl. Univ. di Monte S. Angelo, Edificio 6, Via Cinthia, I-80126, Napoli, Italy \\
$^{3}$ \quad Scuola Superiore Meridionale, Largo S. Marcellino 10, I-80138, Napoli, Italy \\
$^{4}$ \quad Bogoliubov Laboratory for Theoretical Physics, JINR, 141980 Dubna, Russia \\
$^{5}$ \quad National Research Nuclear University MEPhI (Moscow Engineering Physics Institute),	Kashirskoe highway 31, Moscow, 115409, Russia \\
$^{6}$ \quad Astronomical Observatory, Volgina 7, P.O. Box 74, 11060 Belgrade, Serbia}

\corres{Correspondence: dusborka@vinca.rs}

\abstract{After giving a short overview of previous results on constraining of Extended Gravity by stellar orbits, we discuss the Schwarzschild orbital precession of S2 star assuming the congruence with predictions of General Relativity (GR). At the moment, the S2 star trajectory is remarkably fitted with the first post-Newtonian approximation of GR. In particular, both Keck and VLT (GRAVITY) teams declared that the gravitational redshift near its pericenter passage for the S2 star orbit corresponds to theoretical estimates found with the first post-Newtonian (pN) approximation. In 2020, the GRAVITY Collaboration detected the orbital precession of the S2 star around the supermassive black hole (SMBH) at the Galactic Center and showed that it is close to the GR prediction. Based on this observational fact, we evaluated parameters of the Extended Gravity theories with the Schwarzschild precession of the S2 star. Using the mentioned method, we estimate the orbital precession angles for some Extended Gravity models including power-law $f(R)$, general Yukawa-like corrections, scalar-tensor gravity, and non-local gravity theories formulated in both metric and Palatini formalism. In this consideration we assume that a gravitational field is spherically symmetric, therefore, alternative theories of gravity could be described only with a few parameters. Specifically, considering the orbital precession, we estimate the range of parameters of these Extended Gravity models for which the orbital precession is like in GR. We then compare these results with our previous results, which were obtained by fitting the simulated orbits of S2 star to its observed astrometric positions. In case of power-law $f(R)$, generic Yukawa-like correction, scalar-tensor gravity and non-local gravity theories, we were able to obtain a prograde orbital precession, like in GR. According to these results, the method is a useful tool to evaluate parameters of the gravitational potential at the Galactic Center.}


\begin{document}

\section{Introduction}

Several modified gravity theories have been proposed as possible extensions of Einstein's theory of gravity \cite{fisc99}. Among their main motivations, there is the possibility to explain cosmological and astrophysical data at different scales without introducing dark energy and dark matter, but just taking into account further degrees of freedom of the gravitational field emerging from geometric corrections \cite{capo11b}. They have to explain different observations ranging from Solar system, neutron stars, binary pulsars, spiral and eliptical galaxies, clusters of galaxies up to the large-scale structure of the Universe and the observed accelerated dynamics \cite{noji11,noji17,salu21,capo12a}. A number of alternative gravity theories have been proposed (see e.g. \cite{cope04,clif06,clif12,capo11a,capo11b} for reviews). In last years, theories of "massive gravity" have also attracted a lot of attention (see e.g.\cite{ruba08,babi10} and references therein and citations of these papers), especially after a creation of massive gravity models without pathologies such Boulware - Deser ghosts \cite{rham10,rham11,rham14,rham17}. Different alternative approaches for the weak field limit starting from fourth-order theories of gravity, like $f(R)$, have been proposed and discussed \cite{zakh06,zakh07,frig07,nuci07,zakh09,capo09a,iori10,bork12,doku15,doku15b}. The alternative theories of gravity were discussed from the field-theoretical point of view by Petrov et al. in the book \cite{petr17}. Also, some experimental limits related to Extended Theories of Gravity are reported in literature \cite{avil12,duns16,capo13c,capo19}.

In this discussion, stellar dynamics acquires a special role because it allows to investigate gravitational potentials of self-gravitating structures considering stellar motions. In particular, S-stars are the bright stars which move around the Galactic Center \cite{ghez00,scho02,gill09a,gill09b,ghez08,genz10,meye12,gill17,hees17,chu17,amor19,hees20,abut20} where Sgr A$^\ast$ is located which is a compact bright radio source. One of the brightest among S stars with a high eccentricity is called S2 \cite{gill09a,meye12}.

The conventional model for the Galactic Center consists of the supermassive black hole with mass around $4.3\times 10^6 M_\odot$ and an extended mass distribution formed with stellar cluster and dark matter. A total mass of bulk distribution inside a spherical shell where trajectories of bright stars are located must be much smaller than the black hole mass, otherwise, theoretical fits significantly differ from observational data. Recently, GRAVITY Collaboration \cite{abut18,abut19} and the Keck group \cite{do19} evaluated relativistic redshifts of spectral lines for the S2 star near its periapsis passage in May 2018, and the redshifts were consistent with theoretical estimates done in the first post-Newtonian correction for the redshifts.

Some time ago, Ruffini, Arg\"uelles Rueda, \cite{ruff15} proposed a dark matter distribution having a dense core and a diluted halo. Later, the dark matter distribution was called the RAR-model. Recently, Becerra-Vergara et al. \cite{bece21} claimed that this model provides a better fit of trajectories of bright stars in comparison with the supermassive black hole model. The properties of bright star trajectories in gravitational field of a dense core in the RAR-model have been considered in \cite{zakh21} and it was concluded that the gravitational field determined from the RAR model is the harmonic oscillator potential. In this case trajectories of stars are ellipses with centers at the origin and their periods are the same, so they do not depend on semi-major axes, therefore, it was concluded that these properties are not consistent with existing observational data.

In the case of small velocities of stars (in units of speed of light) and great radial coordinates (in Schwarzschild radius units) the orbital precession occurs due to relativistic effects and due to extended mass distribution because we have perturbations of the Newtonian potential in both cases. In the first case precession resulting in a prograde pericentre shift, and in the second case in a retrograde shift \cite{rubi01}, respectively. In both cases result will be rosette shaped orbits \cite{adki07}. Weinberg et al. \cite{wein05} demonstrated that the lowest order relativistic effects, i.e. the prograde precession, could be detectable if the astrometric precision would reach a few tenths of mas.

In our previous papers we constrained different Extended Gravity theories using astronomical data for the S2 orbit \cite{bork12,bork13,zakh14,bork16,zakh16,zakh18,zakh18a,dial19,bork19b,jova21}, fundamental plane of elliptical galaxies \cite{bork16a,capo20,bork21} and barionic Tully-Fischer relation of spiral galaxies \cite{capo17}. In this paper we evaluate parameters of the following theories of gravity: power-law $f(R)$, Yukawa, Sanders, Palatini and non-local gravity by Schwarzschild precession of S2 star. We assume that the orbital precession of the S2 star is equal to the GR prediction $0^\circ.18$ per orbital period. We use this assumption because the GRAVITY Collaboration detected the orbital precession of the S2 star around the SMBH \cite{abut20} and showed that it is close to the corresponding prediction of GR.

It is important to investigate gravity in the vicinity of very massive compact objects because the environment around these objects is drastically different from that in the Solar System framework. The precession of the S2 star is a unique opportunity to test gravity at the sub-parsec scale of a few thousand AU. Gravity is relatively well constrained at short ranges (especially at sub-mm scale) by experimental tests \cite{adel09}, however for long ranges (e.g. at sub-parsec scale) further tests are still needed (see Figures 9 and 10 from \cite{adel09} for different ranges). It is worth stressing that a phenomenological approach can be useful in this context. The precession of the S2 star is an excellent opportunity to test theories of gravity. Also, we use the Schwarzschild precession because it is a new observational result for our Galactic Center and we are interested to learn alternative theory parameters which correspond to the Schwarzschild precession for S2 star.

We organized the paper in the following way. In Section 2 we presented basic theoretical models for power-law $f(R)$, Yukawa, Sanders, Palatini and non-local gravity theories. In Section 3 we evaluate parameters of these theories of gravity by Schwarzschild precession of S2 star and discussed the obtained results. The concluding remarks are given in Section 4.

\section{Theory}

Let us summarize the theoretical models which we want to investigate by S2 star orbits moving around Sgr A$^\ast$.

\subsection{Power-law $f(R)$ gravity}

$f(R)$ gravity is a straightforward extension of GR where, instead of the Hilbert-Einstein action, linear in the Ricci scalar $R$, one considers a generic function of it \cite{clif05,capo06,zakh06,zakh07,capo07,frig07,soti10}. A power-law $f(R)$ is obtained by replacing the scalar curvature $R$ with $f(R)$ = $f_{0n} R^n$ in the gravity Lagrangian \cite{capo06,capo07}. Clearly, for $n=1$, GR is recovered. It is interesting to stress that $n$ can be any positive real number so that, as soon as $n\simeq 1+\epsilon$, with $\epsilon \ll 1$, little deviations from GR can be considered. In the weak field limit, a gravitational potential of the form 	

\begin{equation}
\Phi \left( r \right) = -\dfrac{GM}{2r}\left[ {1 + \left( {\dfrac{r}{r_c}} \right)^\beta} \right],
\label{equ01}
\end{equation}
can be found  \cite{capo06,capo07}. Here $r_c$ is the scale-length parameter and it is related to the boundary conditions and the mass of the system and $\beta$ is a universal parameter related to the power $n$. It is possible to demonstrate that the relation 

\begin{equation}
\beta = \dfrac{12 n^2 - 7n - 1 - \sqrt{36 n^4 + 12 n^3 - 83 n^2 + 50n + 1} }{6 n^2 - 4n + 2}.
\label{equ02}
\end{equation}
holds \cite{capo07}. For the case $n$ = 1 and $\beta$ = 0 the Newtonian potential is recovered.  

Being $n$ any positive real number, it is always possible to recast the $f(R)$ power-law function as

\begin{equation} 
f(R)\propto R^{1+\epsilon}\,.
\label{equ03}
\end{equation}

If we assume small deviation with respect to GR, that is $|\epsilon| \ll 1$, it is possible to re-write  a first-order Taylor expansion as

\begin{equation}
R^{1+\epsilon} \simeq R+\epsilon R {\rm log}R +O (\epsilon^2)\,.
\label{equ04}
\end{equation}

In this way, one can control the magnitude of the corrections with respect to the  Einstein gravity.

\subsection{General Yukawa-like corrections}
	 
Yukawa-like potentials deviate from the standard Newtonian  gravitational potential due to the presence of decreasing exponential terms \cite{talm88,sere06,card11,sand84,iori07,iori08}. Adelberger et al. \cite{adel09} reviewed experiments and constraints on the Yukawa term for the short ranges. In the case of the longer distances parameters of Yukawa gravity potential are given for clusters of galaxies \cite{capo07b,capo09b} and for rotation curves of spiral galaxies \cite{card11}. Other studies of long-range Yukawa-like modifications of gravity can be found in \cite{whit01,amen04,reyn05,seal05,sere06,moff05,moff06}. It is worth noticing that Yukawa-like corrections naturally emerge in the weak field limit derived by  analytic Taylor expansions of $f(R)$ gravity

\begin{equation}
f(R)\,=\,\sum_{n\,=\,0}^{\infty}\frac{f^{(n)}(0)}{n!}\,R^n\, = \, f_0+f_1R+\frac{f_2}{2}R^2+...
\label{equ05}
\end{equation}
Specifically, expanding  with respect to  $R\,=\,0$ that is, around  the Minkowskian background \cite{stab13}, 
it is possible to obtain \cite{sand84,capo11b}:
	
\begin{equation}
\Phi \left( r \right) = -\dfrac{GM}{(1+\delta)r}\left[ {1 + \delta e^{- \left(\dfrac{r}{\Lambda} \right)}} \right],
\label{equ06}
\end{equation}
	
\noindent where $\Lambda$ is the range of interaction  and $\delta$ is a universal constant. In the case $\delta = 0$ the Yukawa potential reduces to the Newtonian one.
	
The Yukawa-like correction coming from $f(R)$ gravity can be used to fix the coefficients in the expansion (\ref{equ05}) and, eventually, to match the observations \cite{capo12}. For the expansion up to the second order, we have 2 parameters to fix \cite{capo14}. The relations between $f_1$, $f_2$ and $\delta$ and $\Lambda$ parameters are $f_1=1+\delta$, $f_2=-(1+\delta)/(\Lambda^2)$ \cite{capo14}.

\subsection{Scalar-tensor gravity and Sanders potential}

The above approach can be improved by adding a scalar field into dynamics in order to match the observations according to the Sanders prescriptions for flat rotation curves of galaxies (see \cite{stab13} and references therein). We can start from the action

\begin{equation}
\mathcal{A}=\int d^{4}x\sqrt{-g}\biggl[f(R,\phi)+\omega(\phi)\,\phi_{;\alpha}\,\phi^{;\alpha}+\mathcal{X}\mathcal{L}_m\biggr]
\label{equ07}
\end{equation}
where $\phi$ is the scalar field. It can be approximated as $\phi\,=\,\phi^{(0)}\,+\,\phi^{(1)}\,+\,\phi^{(2)}\,+\dots$ (see \cite{stab13} for more details) and the function $f(R,\phi)$ with its partial derivatives ($f_R$, $f_{RR}$, $f_{\phi}$, $f_{\phi\phi}$ and $f_{\phi R}$) and $\omega(\phi)$ can be substituted by their corresponding Taylor series. So, in the case of $f(R,\phi)$, we obtain \cite{stab13}:

\begin{equation}
f(R,\phi)\sim f(0,\phi^{(0)})+f_R(0,\phi^{(0)})R^{(1)}+f_\phi(0,\phi^{(0)})\phi^{(1)}...
\label{equ08}
\end{equation}
where Ricci scalar $R$ is approximated as $R\,=\,R^{(0)}\,+\,R^{(1)}\,+\,R^{(2)}\,+\dots$ (see \cite{stab13} for more details). The lowest order of field equations gives \cite{stab13}:

\begin{equation}
f(0,\,\phi^{(0)}) = 0, \quad \quad f_{\phi}(0,\phi^{(0)}) = 0.
\label{equ09}
\end{equation}

In order to evaluate parameters of Sanders gravity we can set the value of the derivatives of the Taylor expansion as $f_{R\phi}\,=\,1,\,\,f_{RR}\,=\,0,\,\,f_R\,=\,\phi$ without losing generality (see \cite{stab13} for more details). Setting the gravitational constant as:
\begin{equation}
G\,=\,\left(\frac{2\,\omega(\phi^{(0)})\,\phi^{(0)}-4}{2\,\omega(\phi^{(0)})\,\phi^{(0)}-3}\right)\frac{G_\infty}{\phi^{(0)}}
\label{equ10}
\end{equation}
where $G_\infty$ is the gravitational constant as measured at infinity and by imposing $\alpha^{-1}\,=\,3-2\,\omega(\phi^{(0)})\,\phi^{(0)}$, the Yukawa gravity potential in the weak field limit gets the following form \cite{stab13}:

\begin{equation}
\Phi_{ST}(\textbf{x})\,=\,-\frac{G_\infty M}{|\textbf{x}|}\biggl\{1+\alpha\,e^{-\sqrt{1-3\alpha}\,m_\phi |\textbf{x}|}\biggr\}.
\label{equ11}
\end{equation}	

The obtained form represents a Sanders-like gravity potential \cite{sand84,sand90} adopted to fit flat rotation curves of spiral galaxies.
It can be written in the folloving form \cite{capo14}:

\begin{equation}
\Phi_{ST}(r)\,=\,-\frac{G M}{(1+\alpha)r}\biggl\{1+\alpha\,e^{-\sqrt{1-3\alpha}\,m_\phi r}\biggr\}.
\label{equ12}
\end{equation}

\subsection{Modified Palatini gravity model}
	
In the Palatini approach, the metric and the connection are considered as independent fields. Metric and Palatini approaches are equivalent in the case of the linear Einstein-Hilbert action, but this is not so for extended gravities, and in particular for $f(R)$ gravity \cite{olmo11}. The Palatini approach leads to second order differential field equations, while the metric approach leads to fourth order coupled differential equations. It is possible also a hybrid metric-Palatini theory, $F(R)=R+f(R)$, where the standard GR term $R$ is metric while the $f(R)$ term is metric-affine. For a review, see \cite{{hark12,capo13,capo13a,capo13b,koiv10,capo12}}. As discussed in \cite{capo13}, it is always possible to reduce a hybrid metric-Palatini model to a scalar-tensor one where the scalar field can be related to the the first derivative of the $f(R)$ term. 

In the weak field limit, the scalar field behaves as $\phi(r) \approx \phi_0 + ( 2G\phi_0 M /3r) e^{-m_\phi r}$, where $M$ is the mass of the system and $r$ is the interaction length. The leading parameters for hybrid Palatini gravity are $m_\phi$ and $\phi_0$ and our aim is to evaluate them. The modified gravitational potential can be written in the following form:

\begin{equation}
\Phi \left( r \right) = -\frac{G}{1+\phi_0}\left[1-\left(\phi_0/3\right)e^{-m_\phi r}\right] M/r\,,
\label{equ13}
\end{equation}
so that it can be considered as a Yukawa-like potential.
	
The parameter $m_{\phi}$ represents a scaling parameter for gravity interaction and $[m_{\phi}]=[Length]^{-1}$ and we measure the parameter in AU$^{-1}$ (AU is the astronomical unit)\footnote{Since dimensions of the S2 star orbit are comparable with the Solar system size and the semi-major axis is around 970 AU, the pericenter distance is around 120 AU.}, $\phi_0$ is dimensionless. Non-vanishing $m_\phi$ and $\phi_0$ would indicate a potential deviation from GR.

\subsection{Non-local gravity model}

Non-local theories of gravity have recently showed that they can suitably represent the behavior of gravitational interaction at both ultraviolet and infrared scales. This behavior could be important because it manages to describe gravitation phenomena without the need for  dark matter and dark energy at different astrophysical scales. The non-locality is introduced by the inverse of the d'Alembert operator. In the references \cite{koiv08,koiv08b,barv14} authors study the dynamics of the non-local theory and its Newtonian limit. See also \cite{capo21,capo20b} for a discussion. 

It is possible to show that the weak-field limit of this theory has the free parameters, $\phi_c, \, r_{\phi}$ and $r_{\xi}$ (see \cite{dial19} for more details). We take specific values for $\phi_c$ = 1 (to obtain the Newtonian limit), and constrain the parameter space of the $r_{\phi}$ and $r_{\xi}$ parameters. The weak field potential for non-local gravity reads:

\begin{eqnarray}
U_{NL} = &-& \frac{G M}{r}\phi_c\ + \frac{G^2 M^2}{2c^2r^2} \left[\frac{14 }{9}\phi _c^2 + \frac{18 r_{\xi }-11 r_{\phi }}{6 r_{\xi} r_{\phi}} r \right] + \nonumber \\
&+& \frac{G ^3 M^3}{2c^4r^3} \left[\frac{7 r_{\phi }-50 r_{\xi }}{12 r_{\xi } r_{\phi }} \phi _c r-\frac{16 \phi_c^3}{27} + \frac{2 r_{\xi }^2-r_{\phi }^2}{r_{\xi }^2 r_{\phi}^2}r^2\right]\, ,
\label{equ14}
\end{eqnarray}
where non-local corrections are evident, where $G$ is the Newtonian constant, $M$ is the black hole mass.

\section{Results and discussion}

In this section we first give a short overview of our previous results regarding the study of the extending theories of gravity by the observed orbit of the S2 star and than we will evaluate parameters of  $R^n$, Yukawa, Sanders, Palatini and non-local gravity theories with the Schwarzschild precession of the S2 star, i.e. by assuming that the  orbital precession of the S2 star is equal to the corresponding theoretical quantity calculated in  GR. In all studied cases we confirmed that these gravity parameters must be very close to those corresponding to the Newtonian limit of the gravity theory. It is rather natural since all bounded orbits of bright stars are very close to ellipses and its foci are located at the Galactic Center, therefore, the gravitational potential should be Newtonian.

In order to calculate the precession of the S2 star in modified gravity, we assume that the weak field potential does not differ significantly from the Newtonian potential. The used perturbing potential is of the form:

\begin{equation}
V(r) = \Phi(r) - {\Phi_N}(r); \quad {{\Phi_N}(r) =  - \dfrac{{GM}}{r}},
\label{equ15}
\end{equation}

\noindent and it can be used for calculating the precession angle according to the equation (30) from paper \cite{adki07}:

\begin{equation}
\Delta \theta = \dfrac{-2L}{GM e^2}\int\limits_{-1}^1 {\dfrac{z \cdot dz}{\sqrt{1 - z^2}}\dfrac{dV(z)}{dz}},
\label{equ16}
\end{equation}

\noindent where $r$ is related to $z$ via: $r = \dfrac{L}{1 + ez}$ and $L = a\left( {1 - {e^2}} \right)$
	
For our calculation we used the results presented in \cite{gill17}, according to which mass of the SMBH of the Milky Way is $M=4.28\times 10^6\ M_\odot$, semi-major axis of the S2 star orbit is $a=0.''1255$, and its eccentricity $e = 0.8839$.

\begin{figure}[ht!]
\centering
\includegraphics[width=0.60\textwidth]{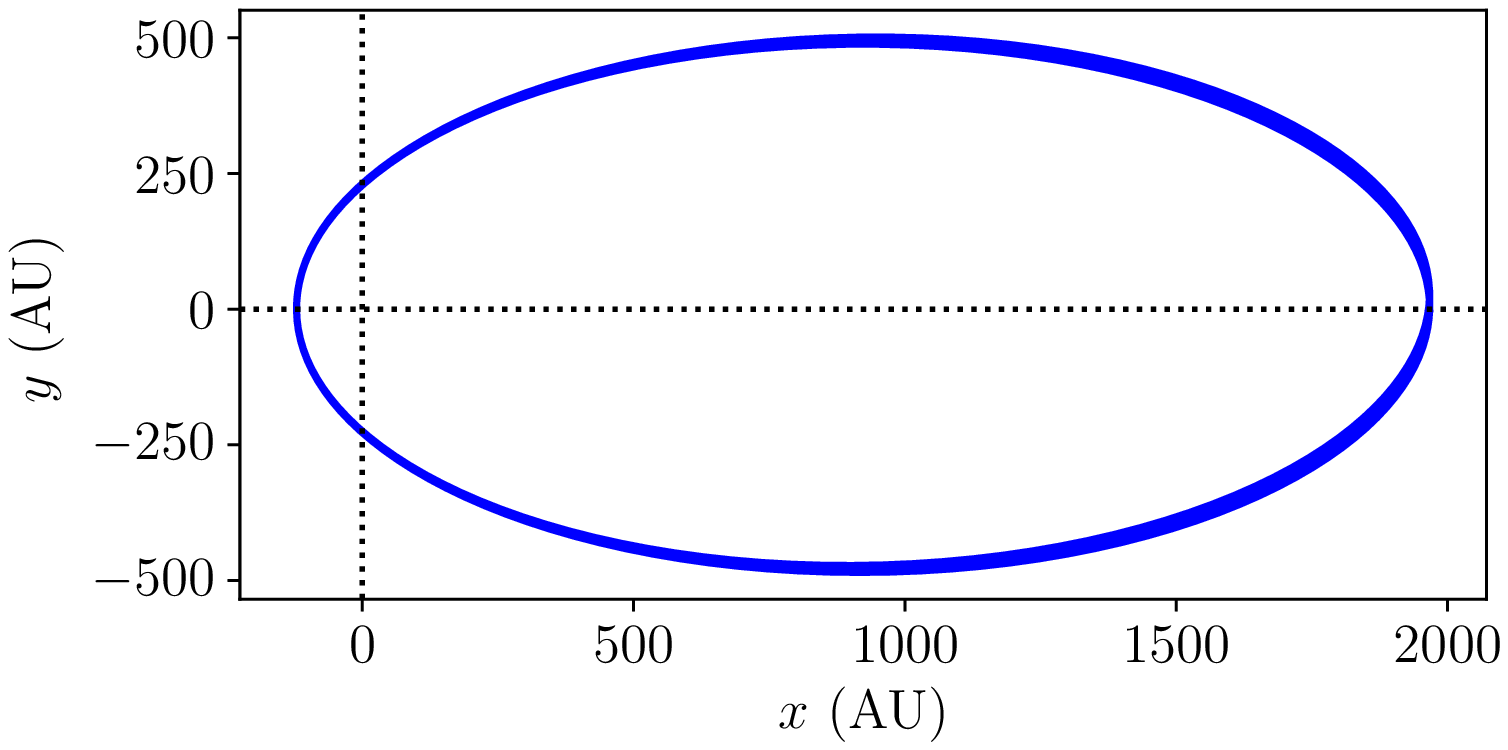} \\
\vspace{0.5cm}
\includegraphics[width=0.60\textwidth]{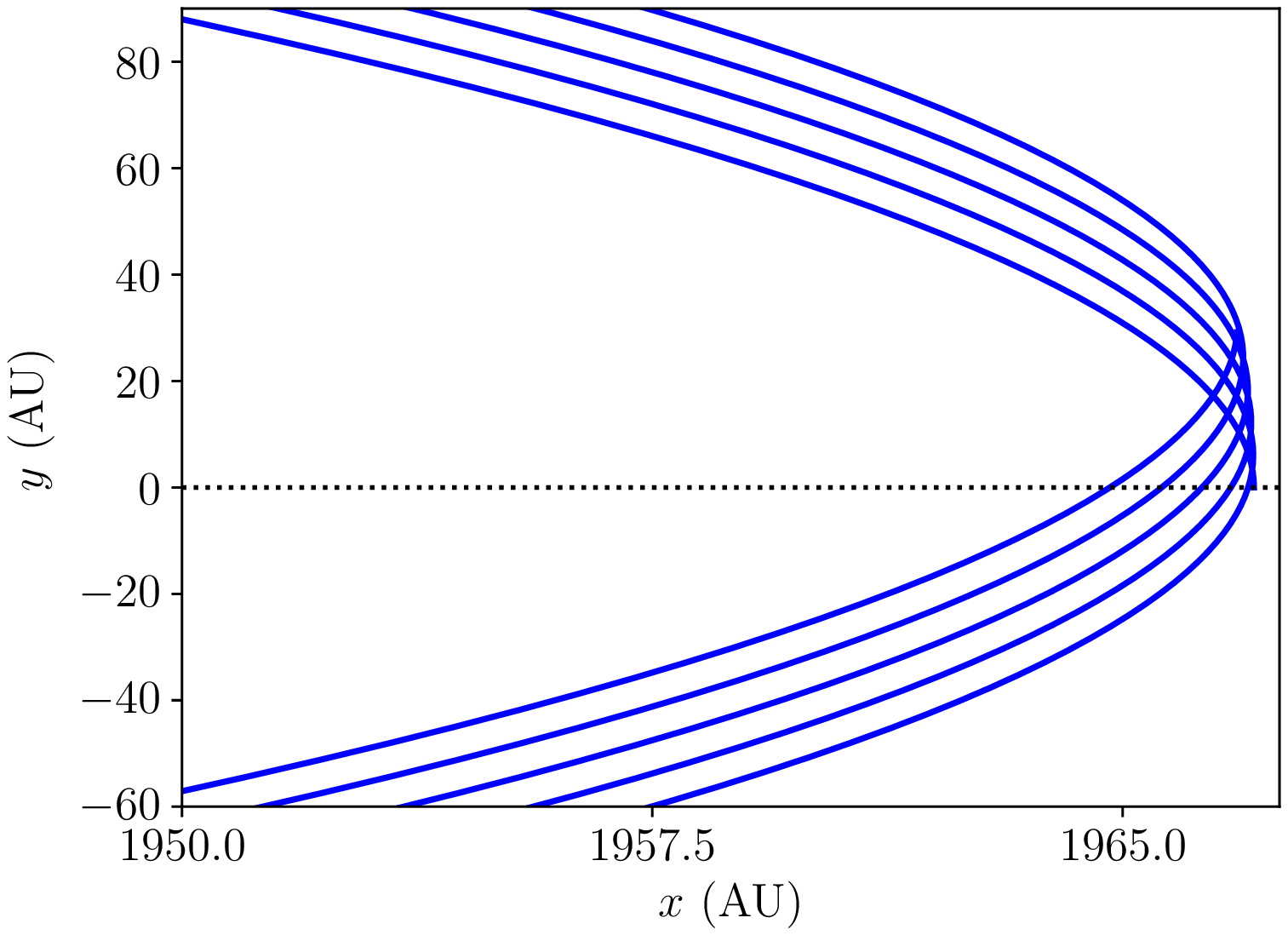}
\caption{The simulated orbits of S2-star in GR. Lower panel represents enlarged part of the upper panel where the obtained values of orbital precession angle is $0^\circ.18$ per orbital period.}
\label{fig01}
\end{figure}

Figure \ref{fig01} shows the simulated orbits of S2-star in GR. Lower panel represents enlarged part of the upper panel. GR predicts that pericenter of S2 star should advance by $0^\circ.18$ per orbital revolution \cite{gill09b}. Since we assume that the orbital precession of the S2 star is equal to the theoretical quantity calculated in GR, the orbits of S2-star will be very close to the orbit in Figure \ref{fig01} for all study cases. We tested the sensitivity and stability of the presented results for Schwarzschild precession of ($0^\circ.18$) by their comparison with the results for slightly different values of Schwarzschild precession: ($0^\circ.178$) and ($0^\circ.182$). We found that the calculations are stable under small perturbations of Schwarzschild precession.

\begin{figure}[ht!]
\centering
\includegraphics[width=0.60\textwidth]{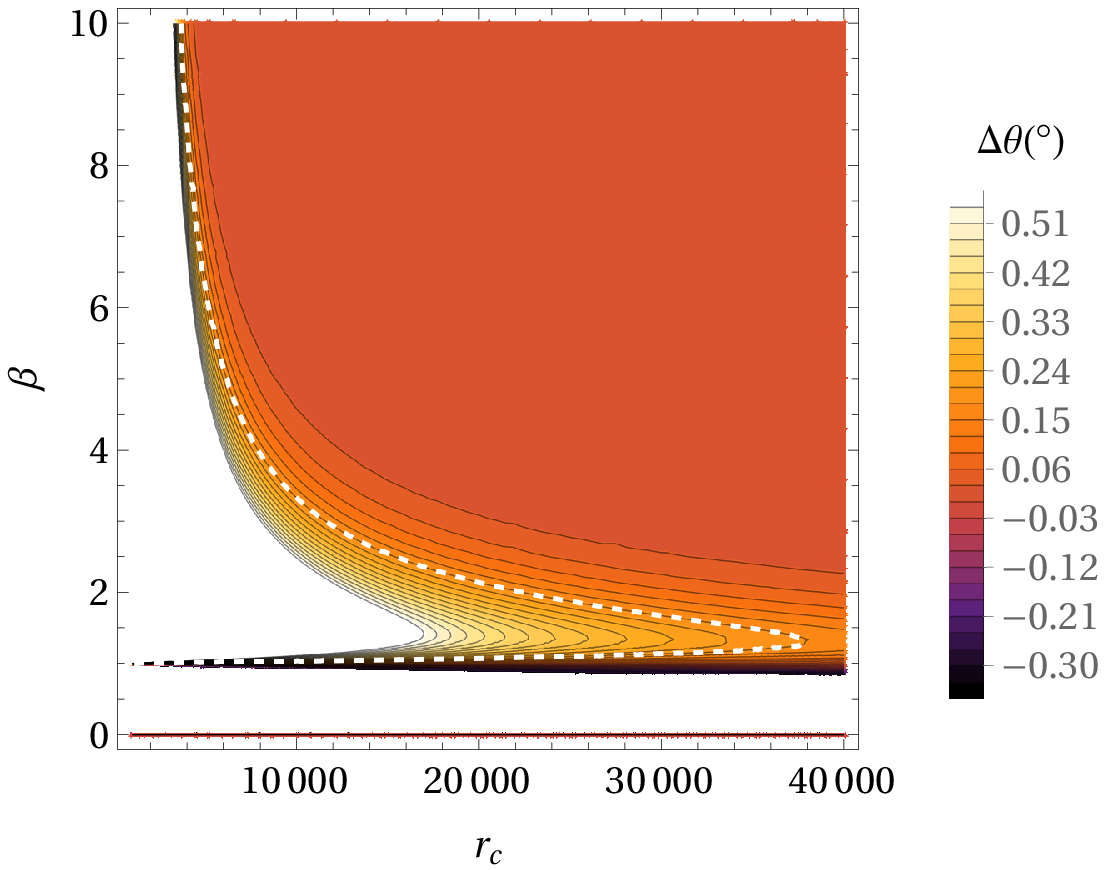} \\
\vspace{0.5cm}
\includegraphics[width=0.60\textwidth]{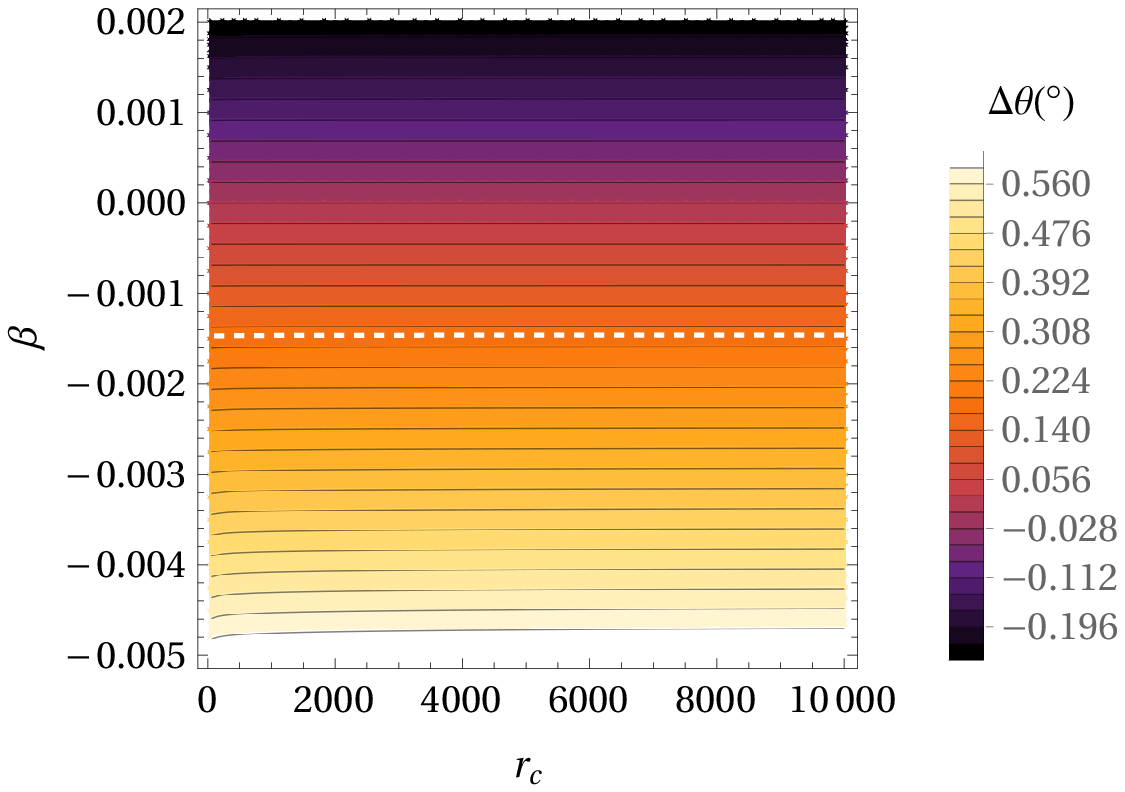}
\caption{The precession per orbital period for $\beta$ - $r_c$ parameter space in the case of $R^n$ gravity potential. With a decreasing value of angle of precession, colors are darker. Locations in parameter space where precession angle has the same value as in GR ($0^\circ.18$) are designated by white dashed line. Lower panel represents the same like the upper panel but for more extended region of $\beta$ - $r_c$ parameter space.}
\label{fig02}
\end{figure}

Figure \ref{fig02} shows the precession per orbital period for $\beta$ - $r_c$ parameter space in the case of $R^n$ gravity potential (\ref{equ01}). Locations in $\beta$ - $r_c$ parameter space where precession angle is the same as in GR are designated by white dashed line. The upper panel shows the results obtained for $\beta > 1$, while the lower panel corresponds to the negative values of $\beta$. In our previous papers regarding $R^n$ gravity \cite{bork12,zakh14} we showed that when parameter $\beta$ was approaching zero, we could recover the Keplerian orbit for S2 star. Our fitting indicated \cite{bork12,zakh14} that optimal value for $\beta$ was around 0.01. The obtained results showed that, in contrast to GR, $R^n$ gravity gives retrograde direction of the precession of the S2 orbit for the values $0<\beta<1$.

However, as it can be seen from the upper panel of Fig. \ref{fig02}, for the values $\beta > 1$ (which are not theoretically justified since $\beta\rightarrow 1$ when $n\rightarrow \infty$, according to the expression (\ref{equ02}), so it can not be incorporated in the theory. In case when parameter $\beta$ is close to -0.0015 (see lower panel of Fig. \ref{fig02}), we can also obtain precession like in GR. This value of  $\beta$ is theoretically justified since it can be obtained from equation (\ref{equ02}) for $n=0.99933$. Regarding $r_c$, it is not possible to obtain the reliable estimate without an additional independent astronomical data set. That is why we believe that $R^n$ gravity could be also considered as a candidate to explain S2-like star orbit, because now findings of the GRAVITY collaboration \cite{abut20} indicate that precession is in the same direction like in GR case.

\begin{figure}[ht!]
\centering
\includegraphics[width=0.60\textwidth]{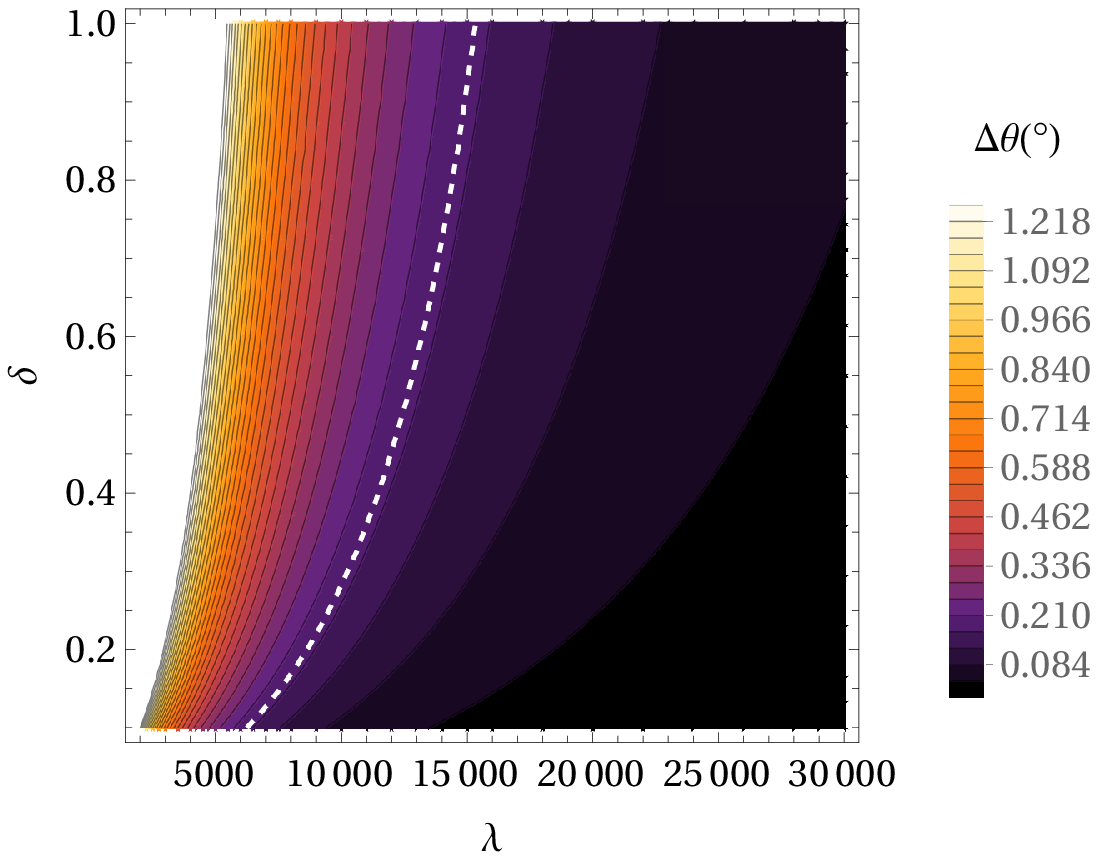} \\
\vspace{0.5cm}
\includegraphics[width=0.60\textwidth]{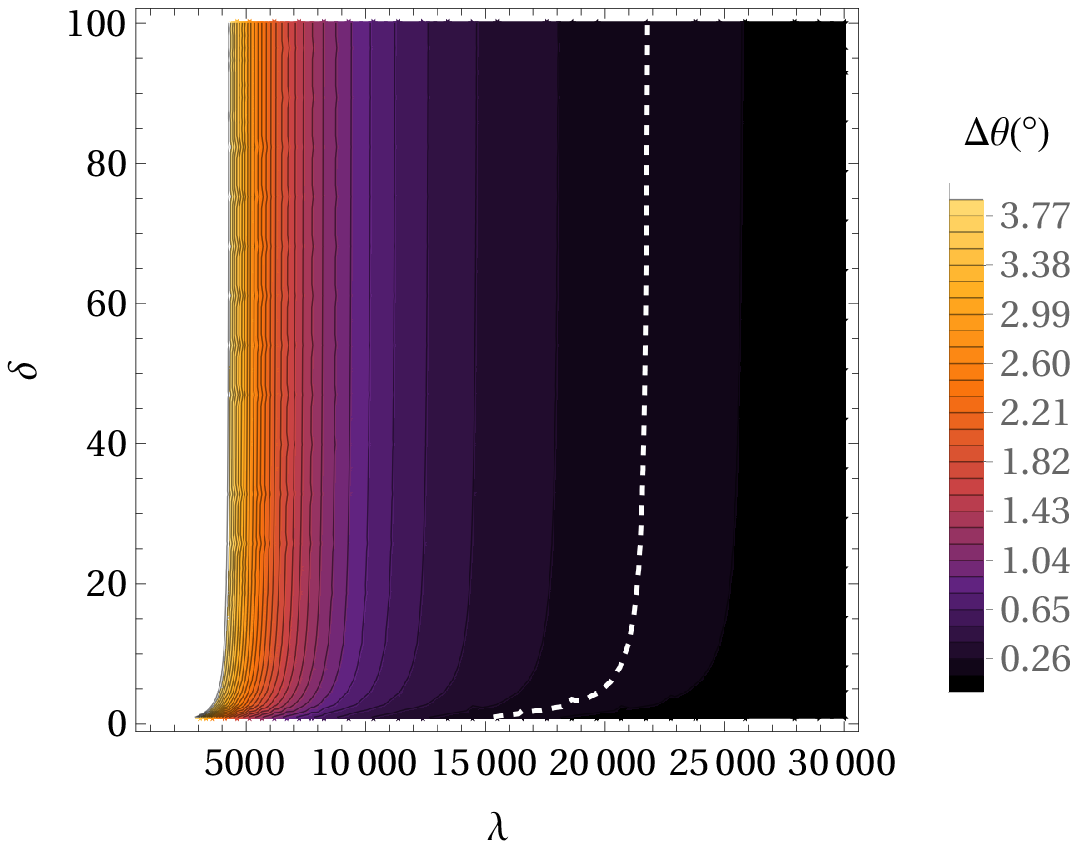}
\caption{The precession per orbital period for $\delta$ - $\Lambda$ parameter space in the case of Yukawa gravity potential. With a decreasing value of angle of precession colors are darker. Locations in parameter space where precession angle has the same value as in GR ($0^\circ.18$) are designated by white dashed line. Lower panel represents the same like the upper panel but for more extended region of $\delta$ - $\Lambda$ parameter space.}
\label{fig03}
\end{figure}

\begin{figure}[ht!]
\centering
\includegraphics[width=0.60\textwidth]{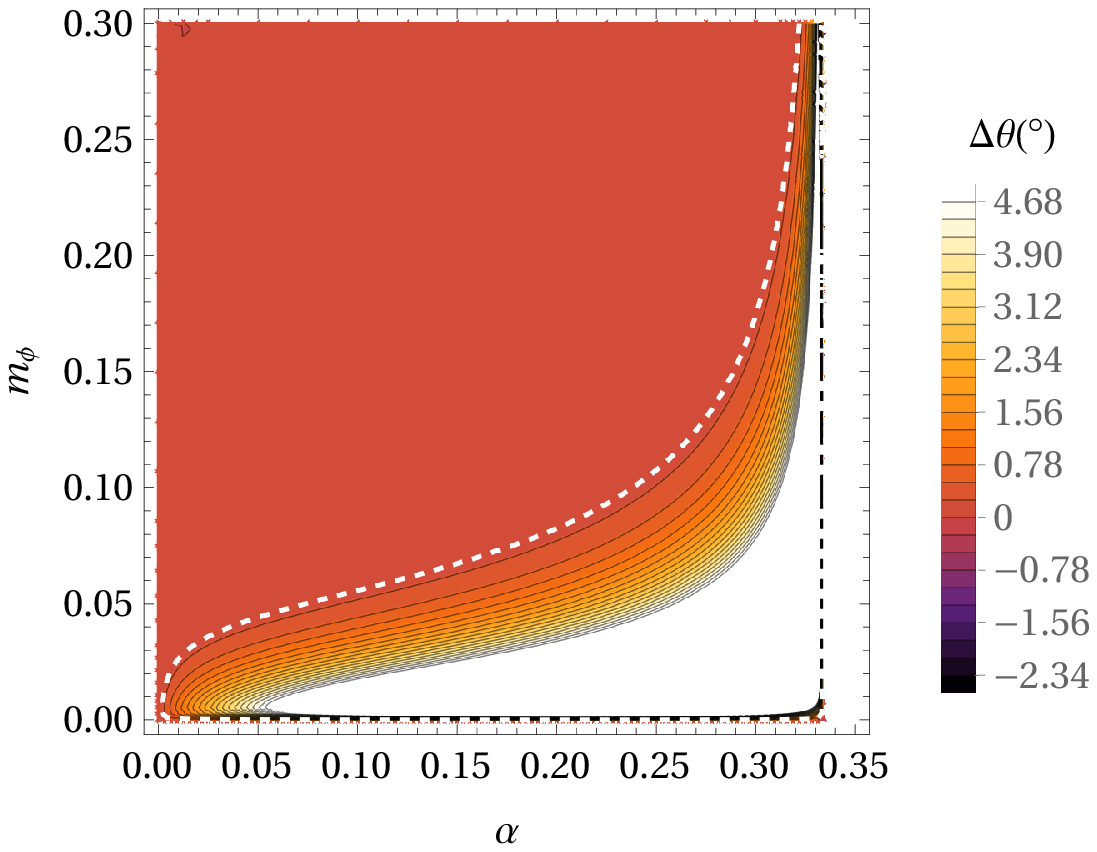} \\
\vspace{0.5cm}
\includegraphics[width=0.60\textwidth]{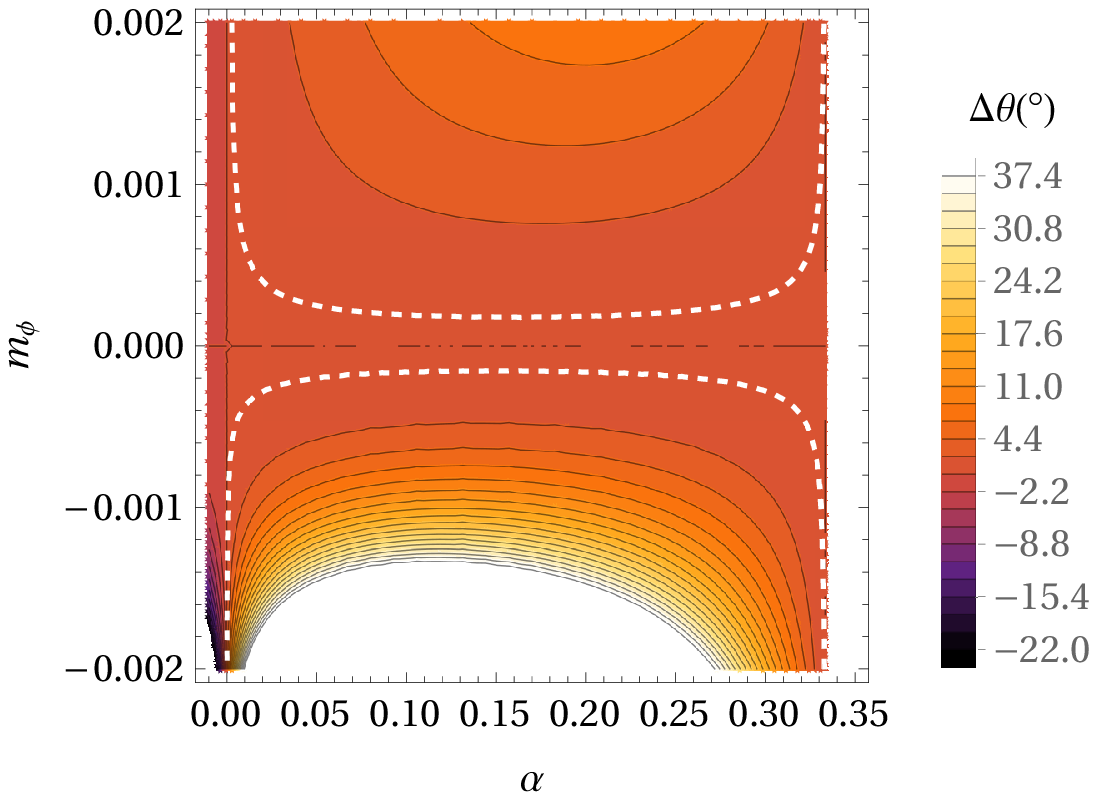}
\caption{The precession per orbital period for $\alpha$ - $m_\phi$ parameter space in the case of Sanders gravity potential. With a decreasing value of angle of precession colors are darker. Locations in parameter space where precession angle has the same value as in GR ($0^\circ.18$) are designated by white dashed line. Lower panel represents the same like the upper panel but for positive values of $m_\phi$ parameter.}
\label{fig04}
\end{figure}

In our previous studies regarding the Yukawa gravity \cite{bork13,capo14,zakh16,zakh18,jova21} we have two parameters to constrain. We used the orbit of the S2 star around the Galactic Centre to test and constrain theory of gravity. In \cite{bork13} we investigated parameters $\delta$ and $\Lambda$, and in \cite{capo14} we investigated values of $f_1$, $f_2$. We find the minimal values of the reduced $\chi^{2}$ in order to estimate $f_1$ and $f_2$ assuming $f_0 = 0$ and in that way we reconstruct $f(R)$ models up to the second order \cite{bork13,capo14}. Figure \ref{fig03} shows the precession per orbital period for $\delta$ - $\Lambda$ parameter space in the case of Yukawa gravity potential. Locations in the $\delta$ - $\Lambda$ parameter space where precession angle is the same as in GR are designated by white dashed line. From our previous findings \cite{bork13,capo14} we conclude that the most probable parameter $\Lambda$ for Yukawa gravity in the case of the S2 star, is around 5000 AU - 7000 AU and that the current observations do not enable us to obtain the reliable constraints on the universal constant $\delta$. Now, with this new strategy we obtained higher value for parameter $\Lambda$, around 21000 AU. For vanishing parameter $\delta$, we recover the Keplerian orbit of S2 star. In the case of the S2 star orbit we obtained the orbital precession in positive direction like in GR. For Yukawa gravity parameters we obtain the following conclusion: it is not possible to get a reliable constraint for $\delta$ based only on the requirement that precession angle is like in GR, and so we need an additional astronomical data set; for $\delta\gg 1$ the obtained values for $\Lambda$ are between 21000 AU and 22000 AU, while for the $\delta\sim 1$, the corresponding values are smaller ($\Lambda\gtrapprox 15000$ AU), as it can be also seen from Table 2 in \cite{zakh18}. If we consider  additional parameters in our model, i.e. that bulk distribution of matter (includes stellar cluster, interstellar gas distribution and dark matter) exists near SMBH in our Galactic Center we will obtain lower value for parameter $\Lambda$ (see paper \cite{jova21} for more details).

We studied Sanders gravity \cite{capo14} and we compared the observed and simulated S2 star orbits around the Galactic Centre in order to constrain the parameters of gravitational potential derived from $f(R,\phi)$ gravity, i.e. Sanders-like potential. We fitted the NTT/VLT astrometric observations of S2 star. The precession of S2 star orbit obtained for the best fit parameter values ($\alpha$ = 0.00018 and $m_\phi$ = -0.0026 AU$^{-1}$) has the positive direction, as in GR. Figure \ref{fig04} shows the precession per orbital period for $\alpha$ - $m_\phi$ parameter space in the case of Sanders gravity potential. Locations in parameter space where precession angle is the same as in GR are designated by white dashed line. In our previous paper \cite{capo14} we obtained much larger orbital precession of S2 star in Sanders-like gravity than the corresponding value predicted by GR. Now, with new strategy we obtained new values for parameters of Sanders gravity: $\alpha$ is between 0 and 1/3 (vertical asymptote in the upper panel of Fig. \ref{fig04} is 1/3, as follows from equation (\ref{equ12})) and $m_\phi$ is between 0 and 0.3 (in AU$^{-1}$ units). From equation (\ref{equ12}) we can see that for $\alpha=0$ or $\alpha=1/3$, or $m_\phi=0$ the Sanders potential is reducing to the Newtonian one. If we want to evaluate $m_\phi$ we need additional independent astronomical data set. As it can be seen from the lower panel of the Fig. \ref{fig04}, for obtaining precession like in GR, absolute value of $m_\phi$ cannot be lower than $\approx$ 0.00065 AU$^{-1}$, which corresponds to $\alpha \approx$ 0.16.

\begin{figure}[ht!]
\centering
\includegraphics[width=0.60\textwidth]{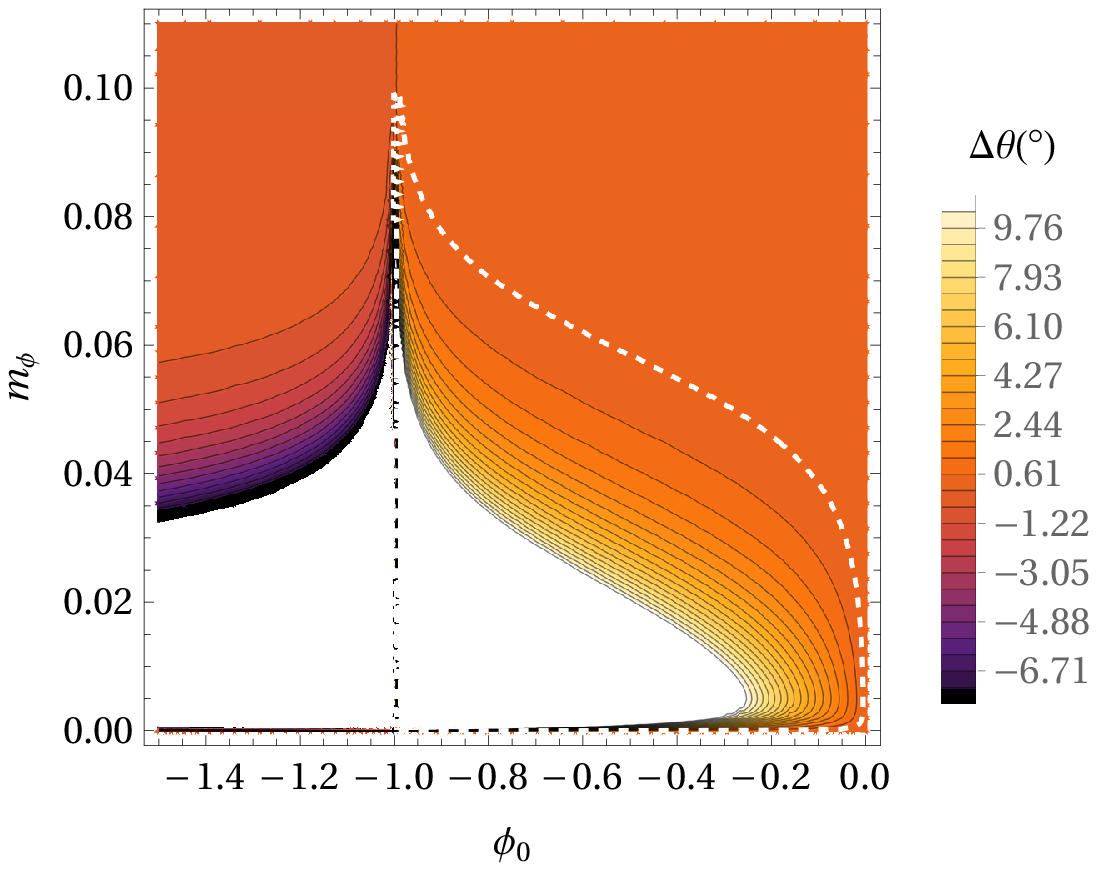} \\
\vspace{0.5cm}
\includegraphics[width=0.60\textwidth]{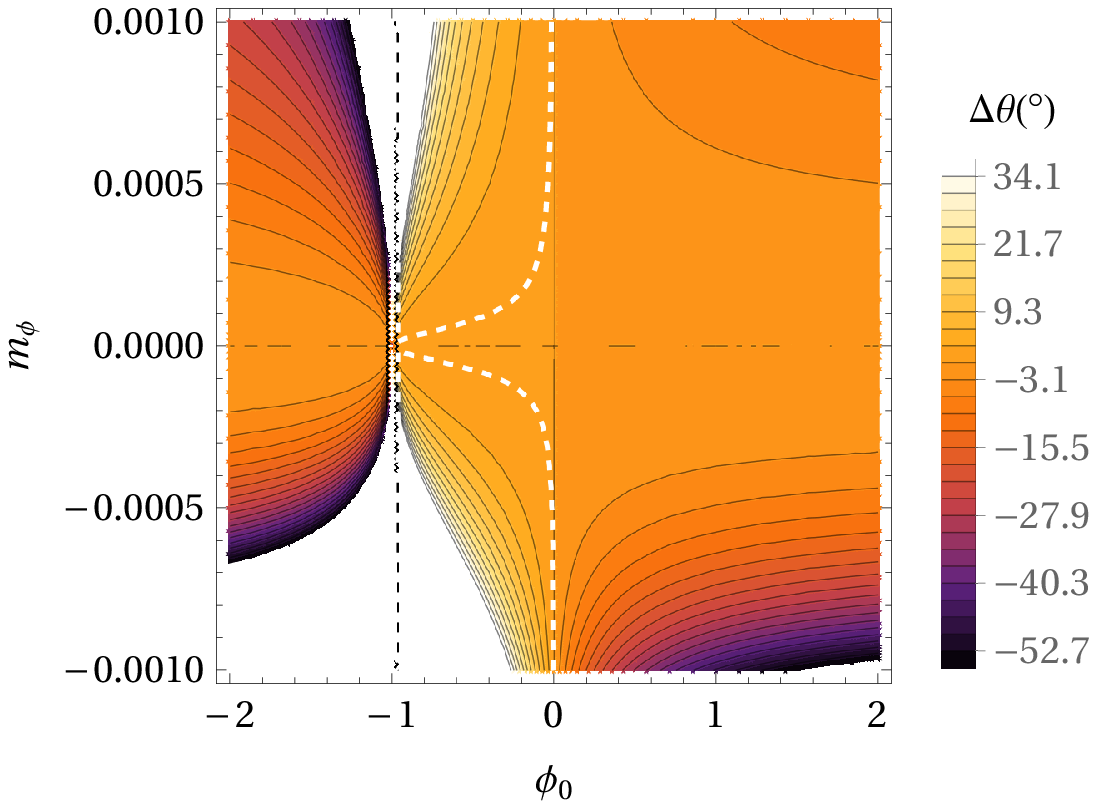}
\caption{The precession per orbital period for $\phi_0$ - $m_\phi$ parameter space in the case of Hybrid Palatini gravity potential. With a decreasing value of angle of precession colors are darker. Locations in parameter space where precession angle has the same value as in GR ($0^\circ.18$) are designated by white dashed line. Lower panel represents the same like the upper panel but for positive values of $m_\phi$ parameter.}
\label{fig05}
\end{figure}

\begin{figure}[ht!]
\centering
\includegraphics[width=0.60\textwidth]{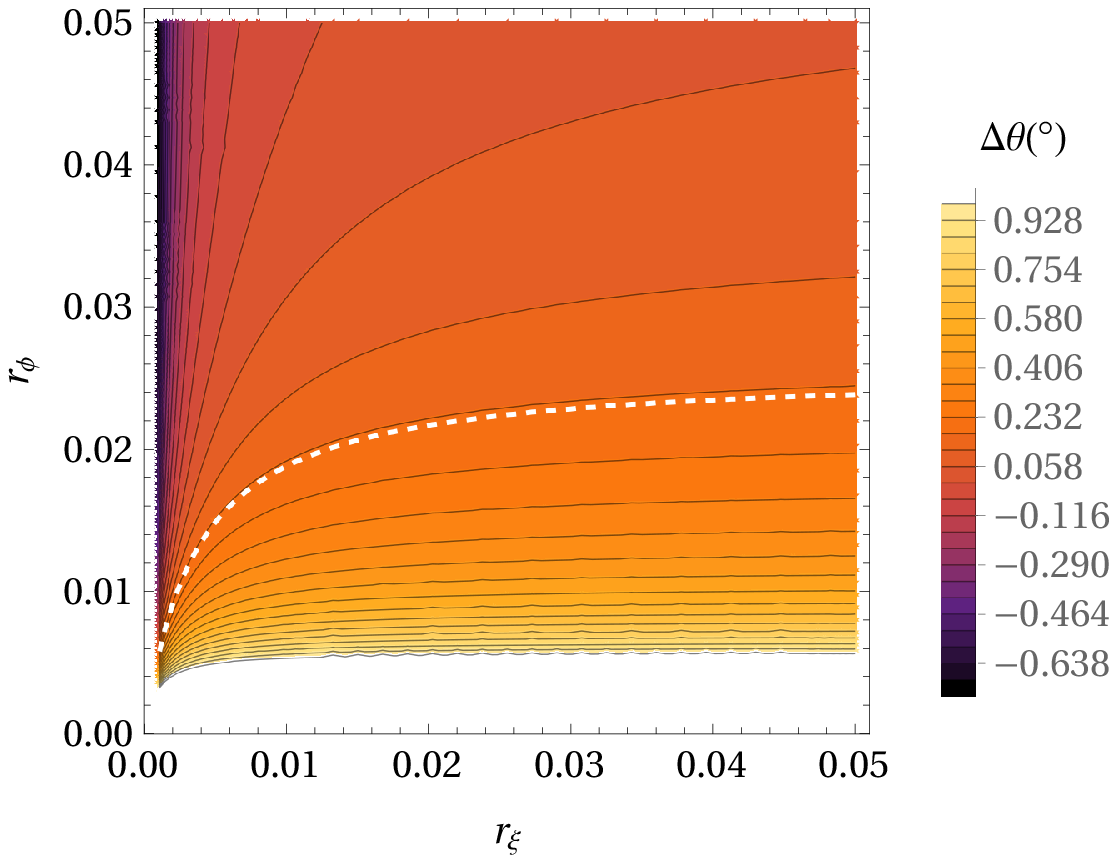} \\
\vspace{0.5cm}
\includegraphics[width=0.60\textwidth]{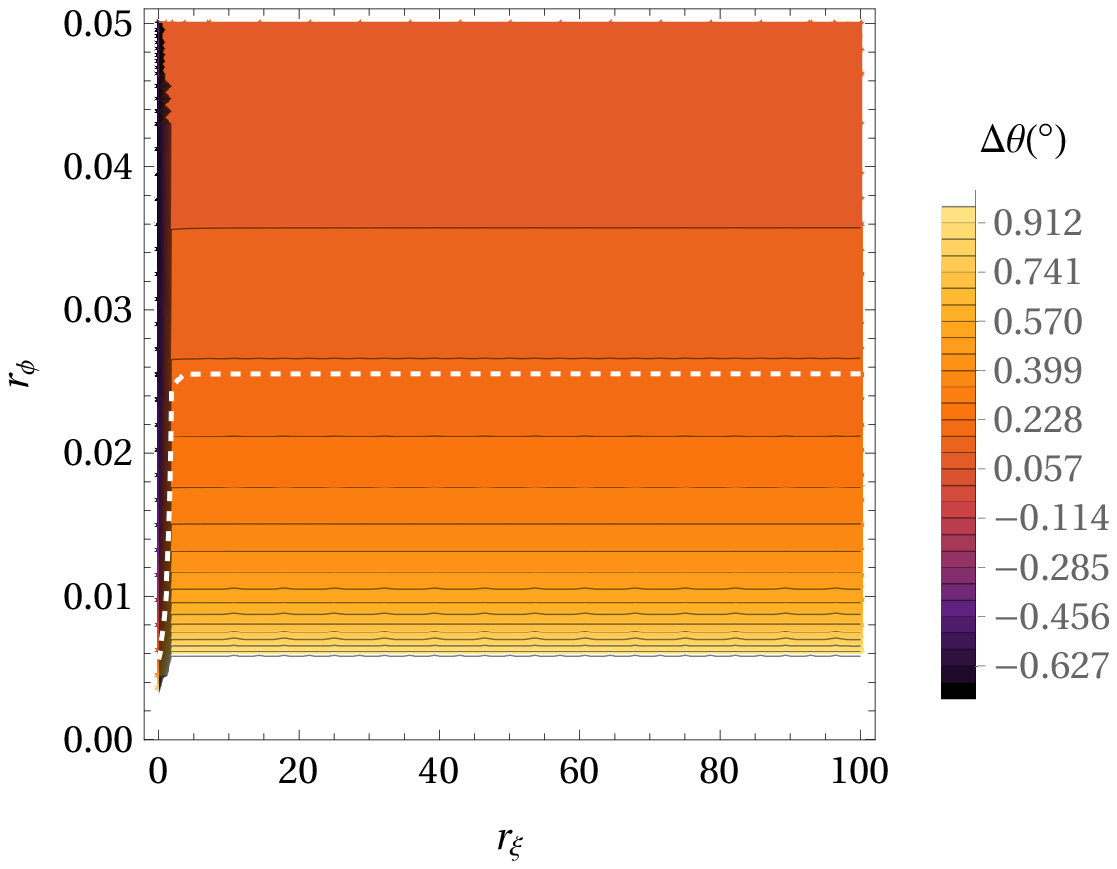}
\caption{The precession per orbital period for $r_{\phi}-r_{\xi}$ parameter space in the case of non-local modified gravity potential. With a decreasing value of angle of precession colors are darker. Locations in parameter space where precession angle has the same value as in GR ($0^\circ.18$) are designated by white dashed line. Lower panel represents the same like the upper panel but for more extended region of $r_{\phi}-r_{\xi}$ parameter space.}
\label{fig06}
\end{figure}

In our previous paper regarding hybrid Palatini gravity \cite{bork16} using the observed positions of S2 star, we constrained the hybrid modified gravity parameters. We obtained following results: the range of values for $\phi_0$ parameter is between -0.0009 and -0.0002; the range of $m_{\phi}$ is between -0.0034 AU$^{-1}$ and -0.0025 AU$^{-1}$; the precession of the S2 star orbit in hybrid Palatini gravity has the same direction as in GR, but the upper limit in magnitude is much bigger than GR. Figure \ref{fig05} shows the precession per orbital period for $\phi_0$ - $m_\phi$ parameter space in the case of hybrid Palatini gravity potential. Locations in $\phi_0$ - $m_\phi$ parameter space where precession angle is the same as in GR are designated by white dashed line. According the obtained figures, we can evaluate parameters of hybrid Palatini gravity. We obtained that $\phi_0$ is between -1 (vertical asymptote in upper panel of Fig. (\ref{fig05})) and 0, which follows from equation (\ref{equ13}) and $m_\phi$ is between -0.1 AU$^{-1}$ and 0.1 AU$^{-1}$. If $\phi_0$ = 0 Palatini potential reduces to Newtonian. Like in the previous case, if we want to evaluate values of $m_\phi$ we need additional independent astronomical data set.	
 
We studied non-local gravity in our previous paper \cite{dial19} and we obtained the values for $r_{\phi}$ and $r_{\xi}$ parameters. We obtained that the most probable value for the scale parameter $r_{\phi}$ is approximately from 0.1 AU to 2.5 AU. We showed that it is not possible to obtain reliable estimates on the parameter $r_{\xi}$ of non-local gravity using only observed astrometric data for S2 star. That length scale is associated with one of the scalar fields which is not dynamical, but only plays an auxiliary role to localize the original non-local Lagrangian. The value of obtained orbital precession of the S2 star has the same order of magnitude as in GR. Figure \ref{fig06} shows the precession per orbital period for $r_{\phi}-r_{\xi}$ parameter space in the case of non-local modified gravity potential. As previously, the locations in $r_{\phi}-r_{\xi}$ parameter space where precession angle is the same as in GR are designated by white dashed line. As it can be seen from the lower panel of the figure \ref{fig06}, for obtaining precession like in GR, $r_{\phi}$ cannot be higher than $\approx$ 0.0256 AU, while, it is not possible to obtain a unique value for $r_{\xi}$ without combinations of these results with some additional astronomical observations.

In this paper we assume that the orbital precession of S2 star is equal to GR value. In our previous papers we fitted astronomical data of the S2 star orbit, and data was with much larger uncertainties. We can notice that results differ from previous one, now it is much more precise. The main reason is that the GRAVITY Collaboration detected the orbital precession of the S2 star and showed that it is close to the GR prediction and direction is the same like in GR.

\paragraph{\textbf{Lense-Thirring precession of the S2 star due to spin of Sgr A*.}}

For the very precise calculations the Lense-Thirring (LT) precession caused by the SMBH spin should be also taken into account in the overall pN precession. In the case of the recently discovered ultra-eccentric S-stars S4711, S62, S4714 \cite{peib20,iori20,iori21,gain20}, the LT precession was studied in \cite{iori20} and it was found that it is much smaller than Schwarzschild precession. Namely, the ratio between the LT precession for a rotating SMBH with dimensionless spin $\chi_g = 0.5$ and the Schwarzschild precession for these stars is around 2.7, 2.3 and 0.86\%, respectively \cite{iori20}.

\begin{figure}[ht!]
\centering
\includegraphics[width=0.65\textwidth]{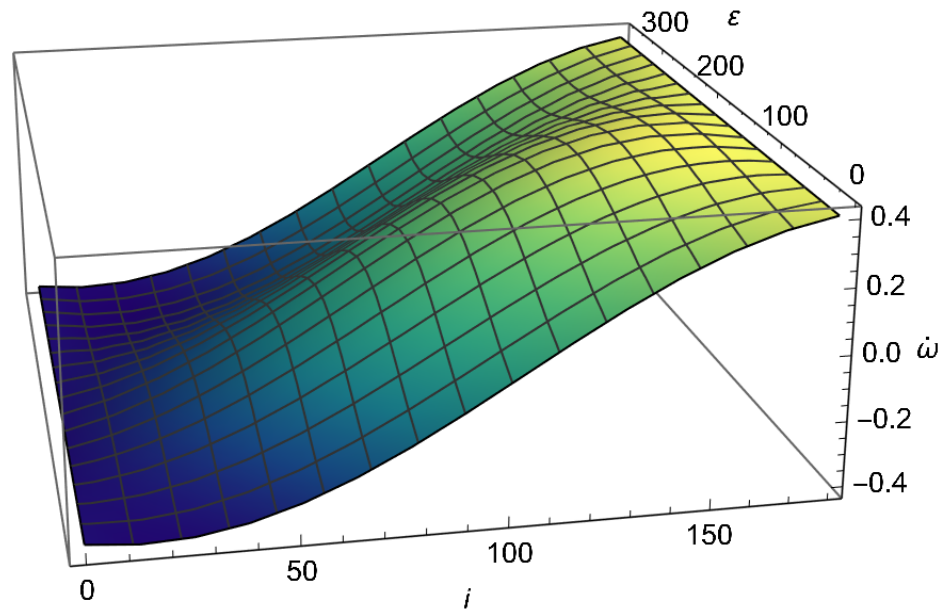}
\caption{The dependence of LT precession $\dot{\omega}\; (''/\mathrm{yr})$ of S2 star on SMBH's spin polar angles $i\; (^\circ)$ and $\varepsilon\; (^\circ)$ for SMBH in Sgr A$^\ast$ with dimensionless spin $\chi_g = 0.5$. SMBH mass is taken from \cite{iori20} and S2 star orbital elements from \cite{abut20}}
\label{fig07}
\end{figure}

In this paper we considered the solutions of alternative theories of gravity for spherically symmetric and static gravitational field, which do not include the SMBH spin. Nevertheless, we estimated the LT precession for S2 star using the expression (10) from \cite{iori20} and assuming a Kerr SMBH rotating with dimensionless spin $\chi_g = 0.5$. We also calculated upper bound for the realistic value of the spin of Sgr A* \cite{frag20}. The spin of Sgr A* was estimated ($\chi_g$ less than 0.1) by the observed distribution of the orbital planes  of the S-stars \cite{frag20}. The obtained dependence of LT precession on SMBH's spin polar angles $i$ and $\varepsilon$ for SMBH spin $\chi_g = 0.5$ is presented in Fig. \ref{fig07}.

The maximum value of LT precession $\dot{\omega} = 0''.423 \; \mathrm{yr}^{-1}$ is obtained for the following polar angles: $i = 160^\circ.8,  \varepsilon = 138^\circ.2$. For the upper bound of the realistic value of the spin of Sgr A*  ($\chi_g = 0.1$) LT precession of $\dot{\omega} = 0''.0845 \; \mathrm{yr}^{-1}$ is obtained for the same values of the polar angles. Taking into account that the Schwarzschild precession of S2 star is equal to $40''.5 \; \mathrm{yr}^{-1}$, it can be seen that the LT precession in the first case is smaller by two orders of magnitude. Therefore, in the case of S2 star the LT precession is much smaller than in the case of the recently discovered ultra eccentric stars S4711, S62 and S4714. Having this in mind, as well as the fact that at the moment the Schwarzschild precession is detected only in the case of S2 star, in our calculations we did not take into account the LT precession.

\section{Conclusions}

In this paper we evaluate parameters of the Extended Gravity theories with Schwarz-schild precession of the S2 star orbit. We assume that the orbital precession of S2 star is equal to the $0^\circ.18$ per orbital period, i.e. the corresponding GR estimate. It is a reasonable approximation since the observed precession angle is close to this quantity \cite{abut20}. We evaluate the parameters of the following extended theories of gravity: $R^n$, Yukawa, Sanders, Palatini and non-local gravity. We can conclude that in all studied cases ($R^n$, Yukawa, Sanders, Palatini and non-local gravity theories), depending on gravity parameters, we can recover prograde orbital precession, like in GR. This is one important condition to evaluate parameters of different modified gravities. Second condition is that value of orbital precession is $0^\circ.18$ per orbital period. Using these two requests we obtained following values of parameters of the extended theories of gravity:

\begin{enumerate}
\item $R^n$: if we take conditions that value of orbital precession is $0^\circ.18$ per orbital period and that it should be prograde like in GR, then $\beta \approx -0.0015$. 
\item Yukawa potential: for $\delta\gg 1$ the obtained values for $\Lambda$ are between 21000 AU and 22000 AU, while for the $\delta\sim 1$, the corresponding values are smaller ($\Lambda\gtrapprox 15000$ AU).
\item Modified Sanders gravity: $\alpha$ is between 0 and 1/3 (vertical asymptote). For $\alpha=0$ or $\alpha=1/3$, or if $m_\phi=0$ the Sanders potential is reducing to the Newtonian one.	
\item Modified Palatini gravity: $\phi_0$ is between -1 (vertical asymptote) and 0. If $\phi_0=0$ the Palatini potential is reducing to the Newtonian one.	
\item Non-local gravity: $r_{\phi}$ is between 0 and 0.0256 AU.
\item For all studied gravity models (described with two parameters) it is not possible to evaluate the both  parameters in a unique way only from conditions that orbital precession is prograde like in GR, and that the precession angle is $0^\circ.18$ per orbital period. In that way, one can use the presented results to find one of the parameters of the studied gravity models, provided that the another one is evaluated in a different way.
\end{enumerate}

If we compare the presented method with our previous studies, it can be seen that now approach is different and obtained values for parameters of gravity models are not the same, but they are the same order of magnitude in both approaches and they are compatible. In our previous papers we fitted astronomical observations of the S2 star orbit, which were obtained with relatively large errors, especially in the first stage of observations (data were collecting for decades). If all orbital parameters are fixed the precision of the S2 star orbit reconstruction is about 1 mas \cite{zakh07}, therefore, an astrometric accuracy must be much better than this value, at least, for one orbital period of S2 star. In this paper we were not fitting observation data, we only assume that the orbital precession of S2 star is equal to the GR value because recently the GRAVITY Collaboration detected the orbital precession of the S2 star and showed that it is close to the GR prediction \cite{abut20}. The Keck collaboration did not declare yet that it detected the Schwarzschild precession for S2 star.

A modified theory of gravity needs to be constrained at different scales: laboratory distances, Solar system, galaxies, binary pulsars, galactic clusters and cosmological scales. Different scales provide constraints at different stringency. The orbit of S2 star is at the sub-parsec scales (a few thousand AU), so we believe that in order to restrict the remaining parameters, one should use additional constraints at similar scales, such as Solar system or binary pulsar observations. In our previous papers we fitted astronomical data of the S2 star orbit and in this paper we assume that the orbital precession of S2 star is equal to GR value because the GRAVITY Collaboration detected the orbital precession of the S2 star and showed that it is close to the GR theoretical estimate. We hope that using this method we can evaluate parameters of alternative models for a gravitational potential at the Galactic Center.

\end{paracol}
\begin{paracol}{2}
\switchcolumn

\authorcontributions{All coauthors participated in writing, calculation and discussion of obtained results.}

\funding{This work is supported by Ministry of Education, Science and Technological Development of the Republic of Serbia. PJ wishes to acknowledge the support by this Ministry through the project contract No. 451-03-9/2021-14/200002. SC acknowledges the support of the Istituto Nazionale di Fisica Nucleare (INFN), sezione di Napoli, iniziative specifiche QGSKY and MOONLIGHT-2.}

\acknowledgments{The authors acknowledge support of Ministry of Education, Science and Technological Development of the Republic of Serbia (DB, VBJ and PJ) and the Istituto Nazionale di Fisica Nucleare (INFN) (SC). DB, VBJ and PJ also wish to thank the Center for mathematical modeling and computer simulations in physics and astrophysics of Vin\v{c}a Institute of Nuclear Sciences.}

\conflictsofinterest{The authors declare no conflict of interest.} 

\abbreviations{Abbreviations}{The following abbreviations are used in this manuscript:\\
	
\noindent 
\begin{tabular}{@{}ll}
GR & General Relativity \\
pN & post-Newtonian \\
SMBH & Super massive black hole \\
\end{tabular}}

\end{paracol}

\reftitle{References}

\end{document}